\documentclass[prb,preprint,groupedaddress,showpacs,showemail]{revtex4}
\usepackage{graphicx}

\begin{document}

\title{Effect of extra electrons on the exchange and  
magnetic anisotropy in the anionic single-molecule 
magnet Mn$_{12}$}
\author{Kyungwha Park$^{1,2}$}\email{park@dave.nrl.navy.mil}
\author{Mark R. Pederson$^{1}$}\email{pederson@dave.nrl.navy.mil}
\affiliation{
$^1$Center for Computational Materials Science, Code 6390,
Naval Research Laboratory, Washington DC 20375 \\
$^2$Department of Physics, Georgetown University, Washington DC 20057}
\date{\today}

\begin{abstract}
To understand the effect of molecular environment on the electronic and 
magnetic properties of the single-molecule magnet (SMM) Mn$_{12}$, 
we explore two possible means for adding extra electrons to molecule.          
We explore both substitution of Mn ions by Fe ions and the inclusion of
neighboring electronic donors. 
For both possibilities we calculate, within density-functional 
theory, the electronic structure, the total ground-state spin and ordering, 
the magnetic anisotropy barrier, the transverse magnetic anisotropy 
parameter $E$ which is responsible for some measured tunneling, and 
the tilting angle of the easy axis from the $z$ axis. Our calculations
show that the total spin increases with increasing number of extra 
electrons except for the case of Mn$_{8}$Fe$_4$ where the resulting
ground state has a low spin. The calculated energy gaps 
between the unoccupied and the occupied orbitals exhibit no clear trend 
as a function of number of extra electrons. The calculated magnetic 
anisotropy barrier decreases with increasing number of extra electrons and
can be directly chased to a quenching of Jahn-Teller distortions at the 
sites where the additional electrons are localized. 
The values of $E$ and the easy-axis tilting angles for the geometries 
with one- and two-extra electrons are significantly larger than those 
induced by solvent disorder.
\end{abstract}

\maketitle


\section{Introduction}

There has been a great demand for high-density magnetic recording 
media that can store a bit of information on a nanoscale magnetic 
particle or molecule. Periodic arrays of high-spin molecules (each
molecule having a volume of a few nm$^3$) could be potentially useful 
for this purpose. These  high-spin molecules consist 
of several transition metal ions which are strongly coupled to each other 
and are surrounded by various ligands. The intermolecular interactions
are quite weak and each molecule has a large energy barrier to magnetization 
reversal (i.e. magnetic anisotropy barrier). This leads to a 
behavior analogous to that of single-domain magnetic nanoparticles 
in an external magnetic field at low temperatures. 
Because of their behavior, the molecules are referred to as 
single-molecule magnets (SMMs). 
Although typical SMMs are paramagnetic, they exhibited magnetic hysteresis 
loops at low temperatures\cite{SESS93,FRIE96} similar to that of 
ferromagnets at room temperatures.  This is because their magnetic moments 
are frozen so that they relax on scales that are slow compared to the 
measurement time. The striking observation on SMMs was that they revealed 
quantum tunneling of magnetic moment\cite{CHUD98} through the 
magnetic anisotropy barrier (MAB), which 
were confirmed by various experiments including temperature independent 
magnetic hysteresis loops\cite{SESS93,FRIE96} and temperature independent 
relaxation time of magnetic moment\cite{PAUL95}. 

Among hundreds of synthesized SMMs, 
[Mn$_{12}$O$_{12}$(CH$_3$COO)$_{16}$(H$_2$O)$_4$]
$\cdot$2(CH$_3$COOH)$\cdot$4(H$_2$O) (hereafter Mn$_{12}$),\cite{LIS80}
has been the most extensively studied experimentally and theoretically.
The strong interest is due to Mn$_{12}$ being the first synthesized SMM 
and to the fact that to date it provides the largest MAB of 
65~K.\cite{BARR97,HILL98,HILL02}
The MAB is directly related to the temperature below which magnetic 
moments are frozen. A general spin Hamiltonian for a SMM is, to lowest order,
\begin{eqnarray}
{\cal H}_0&=& DS_z^2 + E(S_x^2-S_y^2) + g\mu_B \vec{B} \cdot \vec{S}
\label{eq:ham_an}
\end{eqnarray}
where $S_z$ is the spin operator projected onto the $z$ axis which
is the magnetic easy axis, $D$ and $E$ are the uniaxial and transverse 
second-order magnetic anisotropy parameters caused by spin-orbit
coupling. The last term is the Zeeman interaction. In ${\cal H}_{0}$, 
the odd-order terms in $S$ vanish due to time reversal symmetry at 
zero field. High-field, high-frequency electron paramagnetic resonance
(EPR) experiments on the SMM Mn$_{12}$ provided $D=-0.55$~K.
\cite{BARR97,HILL98,HILL02} 
Considering the S$_4$ symmetry of a Mn$_{12}$ molecule (without 
solvent molecules), we know that $E=0$. Quantum tunneling of magnetic 
moment is then allowed for pairs of states that satisfy a certain 
selection rule: The two states must have $S_z$-eigenvalues differing
by integer multiples of 4. However, tunneling between states which 
violate this selection rule are clearly observed in magnetic 
tunneling measurements.\cite{SESS93,FRIE96} To resolve 
this puzzle, three different theories have been proposed.
\cite{CHUD01,CORN02,PARK04-3,PEDE02} 
Chudnovsky and Garanin \cite{CHUD01} proposed any dislocations in single 
crystals could produce a broad continuous distribution in $E$. Detailed 
x-ray diffraction studies by Cornia {\it et al.}\cite{CORN02} suggested 
that disorder caused by acetic-acid (CH$_3$COOH) solvent molecules could 
break the S$_4$ symmetry and provide nonzero locally varying $E$ values. 
Magnetic quantum tunneling\cite{DELB03} and high-field, high-frequency 
EPR experiments\cite{HILL03} showed that their experimental data are not 
in quantitative accord with the Cornia {\it et al.}'s model\cite{CORN02}. 
Recent density-functional 
theory (DFT) calculations\cite{PARK04-3} showed that for quantitative
comparison with experiment, geometry relaxation must be considered. 
Pederson {\it et al.} suggested that coupling 
between the magnetic and vibrational degrees of freedom leads to the
standard fourth-order form of the anisotropy Hamiltonian for 
{\em isotopically pure samples}.\cite{PEDE02} However, 
they showed that for isotopically impure samples\cite{ISOT} 
there are two different vibrational effects that cause symmetry and 
a coupling between $M$ and $M\pm 2$ states and that the symmetry 
breaking would be random assuming the isotopic substitution was.

Very recent experiments on the SMM 
[Mn$_{12}$O$_{12}$(O$_2$CCH$_2$Br)$_{16}$(H$_2$O)$_4$]$\cdot$8CH$_2$Cl$_2$ 
(hereafter Mn$_{12}$-BrAc) showed that there is a broad 
distribution in internal transverse fields (7.3$^{\circ}$ for a
standard deviation).\cite{DELB04}  For this case the molecules of 
crystallization CH$_2$Cl$_2$ have higher symmetry and probably do 
not break the intrinsic S$_4$ symmetry in contrast  to the acetic acid     
solvent molecules in the standard Mn$_{12}$ crystal. It has been suggested 
that the distribution in internal transverse fields could arise from
a distribution in the magnetic easy axes of individual molecules.
\cite{PARK-JCP,DELB04} But the origin of the distribution 
in the magnetic easy axes remains unknown. In this regard, we consider,
within DFT, the possibility that extra electrons when accepted by Mn$_{12}$ 
molecules, will exhibit a disorder- or defect-induced behavior that is
observed in the experimental samples. 

The SMM Mn$_{12}$ has four Mn$^{4+}$ ions at the corners of 
the inner cubane and eight Mn$^{3+}$ ions in the outer crown 
(Fig.~\ref{fig:Mn8Fe4}). The magnetic moments of the eight 
Mn$^{3+}$ ions ($S=2$) were observed to be antiparallel to those of 
the four Mn$^{4+}$ ions ($S=3/2$), which leads to a total ground-state 
spin of $S=10$.\cite{GOTO03} The observed magnetic structure 
and the MAB for the ground state of the Mn$_{12}$ has been calculated
in recent DFT calculations.\cite{PEDE99} The exchange interactions 
between metal ions have also been calculated\cite{PARK04-1} within DFT.
Both the MAB and exchange interactions are 
in very good agreement with experiments.\cite{REGN02} Despite an increasing 
number of synthesized SMMs, it is not well understood what types of variations
in the the molecular environment influence the magnetic 
structure, the intramolecular exchange interactions, and the MAB of 
the molecule. Hendrickson's and Christou's groups synthesized two 
types of anionic SMM Mn$_{12}$ and performed magnetic measurements 
on them.\cite{SCHA94,EPPL95,SOLE03} One type is gleaned by substitution of
four Mn ions by four Fe ions in the ordinary Mn$_{12}$ structure, 
[Mn$_{8}$Fe$_4$O$_{12}$(CH$_3$COO)$_{16}$(H$_2$O)$_4$] (hereafter
experimental Mn$_8$Fe$_4$).\cite{SCHA94} The other is synthesized by 
adding one or two PPh$_4$ molecules (Ph$\equiv$C$_6$H$_5$) to a Mn$_{12}$ 
molecule containing different ligands from the ordinary Mn$_{12}$,
(PPh$_4$)[Mn$_{12}$O$_{12}$(O$_2$CCH$_2$CH$_3$)$_{16}$(H$_2$O)$_4$] 
(hereafter (PPh$_4$)Mn$_{12}$) and
(PPh$_4$)$_2$[Mn$_{12}$O$_{12}$(O$_2$CCHCl$_2$)$_{16}$(H$_2$O)$_4$]
(hereafter (PPh$_4$)$_2$Mn$_{12}$). A
PPh$_4$ molecule has one unpaired electron with A$_1$ symmetry and may
be thought of as a molecular alkali.\cite{EPPL95,SOLE03} 
Magnetic measurements on both types revealed that their total spins, 
magnetic susceptibilities, and MABs significantly 
changed with the number of extra electrons but that the two
types are partially but not entirely homologous.\cite{SCHA94,EPPL95,SOLE03} 
As far as we know, no theoretical studies on these interesting results
have been reported. 

As implied earlier, our motivation is twofold. One is to understand
how the microenvironment within a Mn$_{12}$ molecule controls 
its magnetic properties. The other is to view extra 
electrons as disorder or defects in the Mn$_{12}$ structure and to 
understand their effects on S$_4$ symmetry-breaking terms in
the spin Hamiltonian Eq.~(\ref{eq:ham_an}). 
Motivated by the experiments described in the previous paragraph, 
we employ DFT to investigate two categories of environments where extra 
electrons can be transferred to metal ions in a single Mn$_{12}$ 
molecule: (i) substitution of Fe for Mn and (ii) addition of electronic
donors such as K to the molecule. In the first case, since an Fe atom 
has one more electron than a Mn atom, replacing Mn by Fe will play 
the role of donating one extra electron to the ordinary Mn$_{12}$ structure. 
In the second case, since a PPh$_4$ molecule has one unpaired electron, 
addition of PPh$_4$ molecules to a Mn$_{12}$ molecule is equivalent 
to adding single-electron donors to the Mn$_{12}$ molecule.
In our calculations, K atoms were used instead of large PPh$_4$ molecules 
because a K atom also has one unpaired electron and it reduces computational 
cost. In the sense that there are extra electrons in the ordinary
Mn$_{12}$ structure, we call the SMMs under the new environment 
the anionic SMM Mn$_{12}$. For both categories, we consider one-, two-, 
and four- extra electrons donated to the Mn$_{12}$ molecule. 
For each anionic state, we calculate the electronic structure,
the total magnetic moment, and the magnetic anisotropy parameters, using
DFT. We find that Fe substitution provides similar results (not exactly
same) to K addition and that some microenvironments drastically alter 
the magnetic structure. In Sec.~II, we describe our detailed model 
and method. In Secs.~III and IV, we separately present DFT calculations on 
the electronic and magnetic properties for Fe substituted and K added 
geometries. In Secs.~V and VI, we present our discussion and conclusion.

\section{Model and Method}

Our DFT calculations\cite{KOHN65} are performed with spin-polarized 
all-electron Gaussian-orbital-based Naval Research Laboratory Molecular 
Orbital Library (NRLMOL)\cite{PEDE90}. Here the Perdew-Burke-Ernzerhof 
(PBE) generalized-gradient approximation (GGA) is used for the 
exchange-correlation potential.\cite{PERD96} For all atoms full basis 
sets are used with a fine numerical integration mesh unless 
otherwise specified.\cite{PORE99} Prior to 
geometry relaxation, an initial spin configuration must be assigned 
with an initial total 
moment that can either be fixed or optimized. An optimized geometry is 
obtained when forces on all atoms become small. To find the 
ground-state spin configuration, different spin configurations
with different total magnetic moments are considered. 
The second-order magnetic anisotropy parameters are calculated for
an optimized geometry using the technique 
described in Ref.[\onlinecite{PEDE99}].

As initial geometries for the two types of environmental changes, we 
start with a simplified form of a Mn$_{12}$ molecule with S$_4$ symmetry, 
[Mn$_{12}$O$_{12}$(HCOO)$_{16}$(H$_2$O)$_4$]. Here 16 acetates 
(CH$_3$COO) in the experimental Mn$_{12}$ were replaced by 16 formates (HCOO) 
(refer to Fig.~\ref{fig:Mn8Fe4}). Earlier calculations on 
the Mn$_{12}$ molecule showed that the energy gap between the 
majority HOMO (highest occupied molecular orbital) and the minority 
(majority) LUMO (lowest unoccupied molecular orbital) is about 0.804~eV 
(0.438~eV).\cite{PEDE99} 
The calculated majority HOMO-LUMO energy gap agreed with the result 
of the recent electrical resistivity measurement, 0.37$\pm$0.05~eV.\cite{NORT03} 
The total ground-state moment was calculated to be 20$\mu_B$, which agreed
with experiment. The calculated second-order MAB 
was 54~K in zero field, which did not change much with our simplification 
of the molecule. Hereafter, unless specified, Fe substitution and K 
addition will be carried out in the above simplified form of a Mn$_{12}$ 
molecule.

For the optimized geometry of the Mn$_{12}$ molecule, the first
four unoccupied majority orbitals 
directly above the Fermi level are close to each other in energy but well 
separated from the rest of unoccupied orbitals. The four orbitals contain 
the LUMO, an orbital 0.008~eV above the LUMO, and doubly 
degenerate orbitals 0.012~eV above the LUMO. 
The fifth orbital is 0.212~eV above the LUMO.  Because the 
third and fourth orbitals are degenerate, the case of three extra 
electrons is not considered in our DFT calculations. But our prediction 
of the case based on symmetry will be provided in Sec.~V. 
We first examine whether 
there are any favorable sites for the extra electrons in the Mn$_{12}$
geometry when four Fe ions are substituted in a S$_4$ symmetric fashion. 
Then when four-, two-, and one-extra electrons are accepted in a Mn$_{12}$ 
molecule under each type of environment, we calculate HOMO-LUMO energy gaps, 
the total ground-state moment, and the ground-state spin configuration.
In various ways, we examine whether the extra electrons are well 
localized at some of the Mn sites. Using the optimized geometry with
the ground-state spin configuration, we calculate the MAB, the transverse 
magnetic anisotropy parameter $E$, and the magnetic easy axis. When 
experimental data are available, comparison between calculations and
experiment are discussed.

\section{DFT calculations: Iron substitution}

\begin{figure}
\includegraphics[angle=0,width=0.8\textwidth]{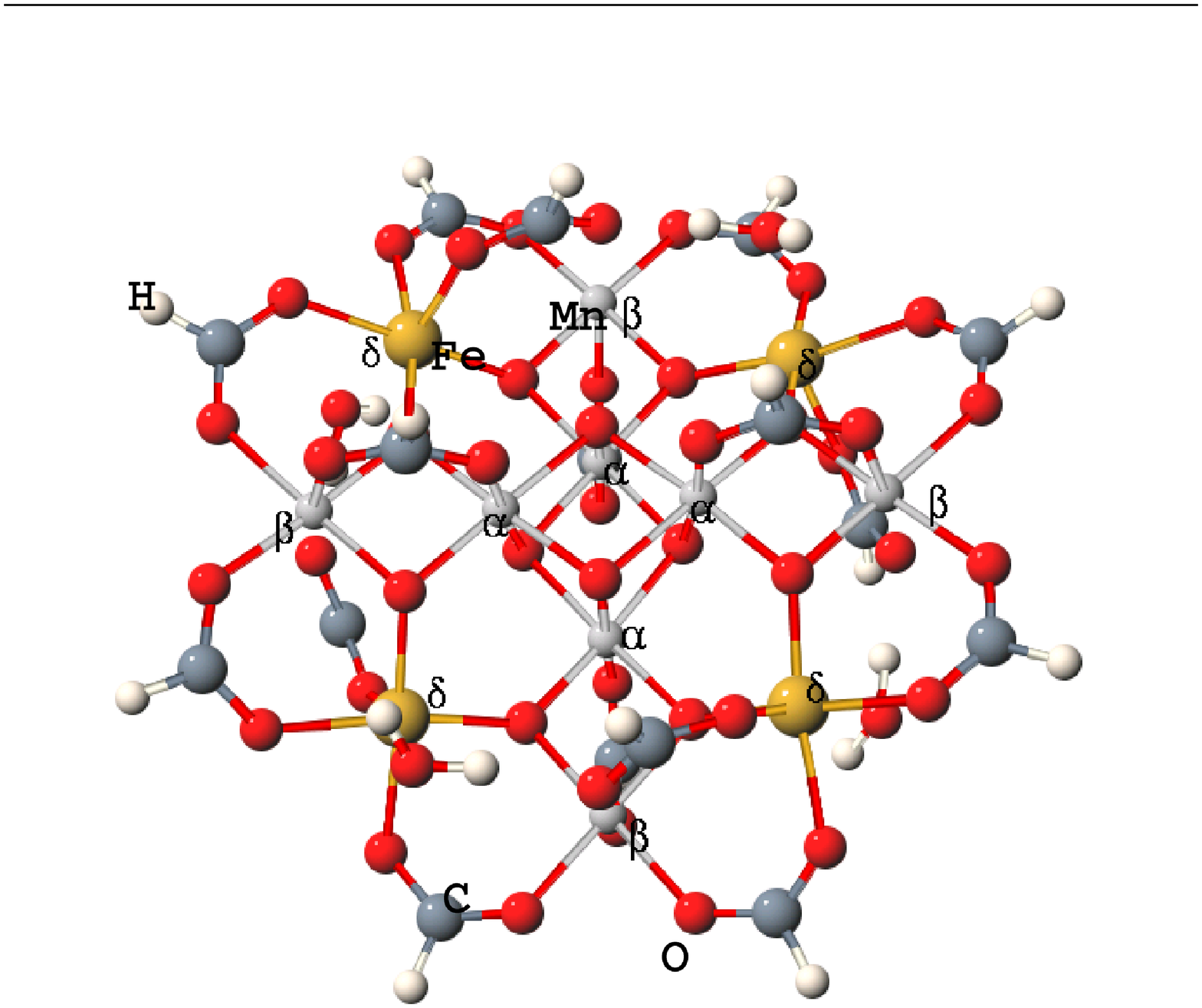}
\includegraphics[angle=0,width=0.35\textwidth]{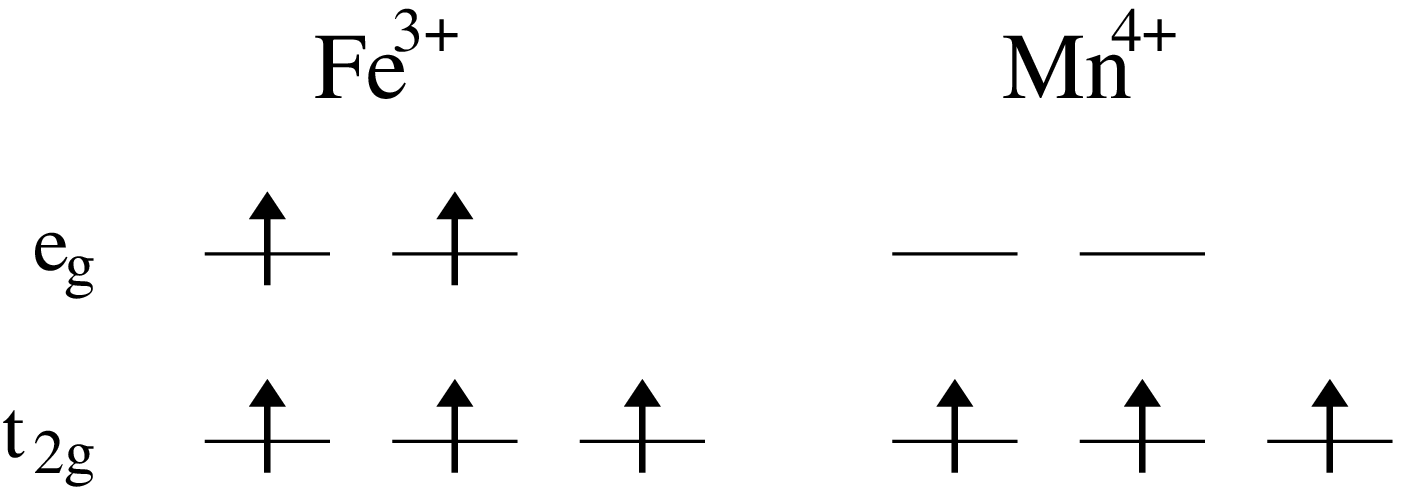}
\caption{Schematic diagram of the fourfold symmetric tetra-Fe
substituted Mn$_{12}$ geometry, Mn$_{8}$Fe$_4$O$_{12}$(HCOO)$_{16}$(H$_2$O)$_4$.
The shading scheme of the atoms is as follows: darkness increases 
from H (white), to Mn, to Fe (largest spheres), to C, and to O (darkest). 
The three inequivalent metal ion sites are labeled as $\alpha$, $\beta$, and 
$\delta$. The bottom figures illustrate occupancy of the $e_g$ and $t_{2g}$ 
3$d$ orbitals for Fe$^{3+}$ (or Mn$^{2+}$) and Mn$^{4+}$.}
\label{fig:Mn8Fe4}
\end{figure}

\begin{figure}
\includegraphics[angle=0,width=0.8\textwidth]{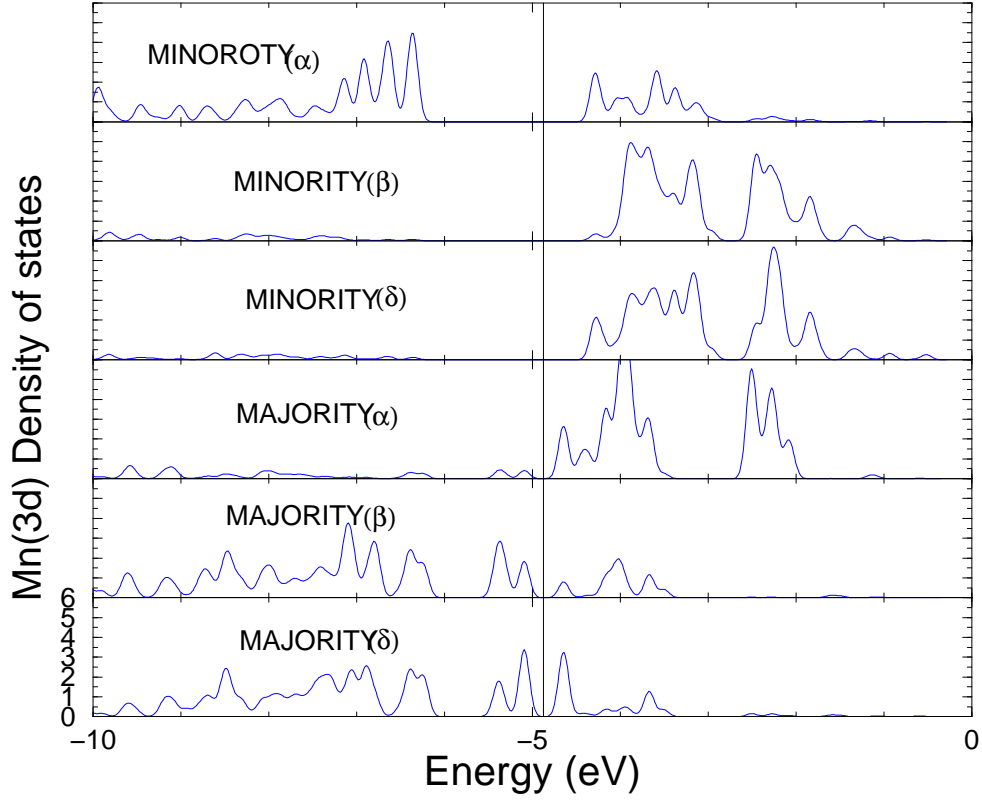}
\caption{Calculated density of states projected onto the majority and minority
Mn($3d$) orbitals at each of the three inequivalent sites ($\alpha$, $\beta$,
and $\delta$) for the fourfold symmetric
[Mn$_{12}$O$_{12}$(HCOO)$_{16}$(H$_2$O)$_4$] whose structure is exactly the
same as Fig.~\ref{fig:Mn8Fe4} with Fe replaced by Mn. The solid vertical line 
denotes the Fermi energy level.}
\label{fig:DOS}
\end{figure}

The S$_4$ symmetry and the geometry of the Mn$_{12}$ provide three 
symmetrically inequivalent Mn sites such as one inner cubane site
($\alpha$) and two inequivalent outer crown sites ($\beta$ and 
$\delta$) shown in Fig.~\ref{fig:Mn8Fe4}. To locate an energetically
favorable site for accepting extra electrons, 
we consider the three S$_4$ symmetric configurations in which four
Mn ions are replaced by Fe ions at the three inequivalent sites, 
[Mn$_{8}$Fe$_4$O$_{12}$(HCOO)$_{16}$(H$_2$O)$_4$] (hereafter Mn$_8$Fe$_4$).
When four Mn ions at the sites $\alpha$ are replaced, 
the configuration is called the geometry $\alpha$. A similar definition is 
applied to the geometries $\beta$ and $\delta$. The geometry 
$\delta$ is shown in Fig.~\ref{fig:Mn8Fe4}. The energies of the three 
tetra-Fe substituted geometries are self-consistently calculated 
with the total magnetic moment fixed and with the spin configuration 
illustrated in Fig.~\ref{fig:spin_config}(a) (only the direction of 
each Mn moment is assigned such as spin up or down). Then we calculate 
HOMO-LUMO energy gaps and check whether Fermi filling is satisfied
for the three geometries with two different fixed moments 
(24$\mu_B$ and 22$\mu_B$). If there exist appreciable HOMO-LUMO energy
gaps and the calculated orbitals abide by the Fermi filling, then
the system is regarded as being magnetically stable. As summarized in 
Table~\ref{table:four_Fe_sub}, the energy is the lowest among the six possible
configurations when four Fe ions occupy the site $\delta$ given the 
total moment of 24$\mu_B$. Local moments around the Fe sites 
for the geometries $\beta$ and $\delta$ with the total moment of
24$\mu_B$, are found to be 3.94$\mu_B$ and 3.95$\mu_B$ within 
a sphere with 2.23 Bohr radius (Table~\ref{table:four_Fe_sub}). These local 
moments are definitely larger than those at the other outer Mn sites, 
3.59-3.64$\mu_B$, for the Mn$_8$Fe$_4$ geometries or those at the 
outer Mn sites, 3.55-3.62$\mu_B$, for the Mn$_{12}$ geometry. But 
the local moment at the Fe site for the geometry $\alpha$ is greatly 
reduced to $-1.55$~$\mu_B$. This is because the initial spin configuration 
[Fig.~\ref{fig:spin_config}(a)] does not permit the extra electrons to be
localized in the cubane. The same trend appears for the local moments of
the geometry with the total moment of 22$\mu_B$. Our finding of the 
$\delta$ to be energetically favored is also evidenced in the site
calculated density of states projected onto the three inequivalent 
Mn($3d$) orbitals for the Mn$_{12}$ geometry. As shown in Fig.~\ref{fig:DOS}, 
the preferred site $\delta$ has the largest majority density of states 
directly above the Fermi level. This finding is also corroborated by x-ray 
crystallographic
data on the experimental Mn$_8$Fe$_4$.\cite{SCHA94} 
Henceforth we consider only cases where Fe is substituted for Mn 
at the site $\delta$.

\subsection{Four Fe substitution: Mn$_8$Fe$_4$}

\begin{figure}
\includegraphics[angle=0,width=0.42\textwidth]{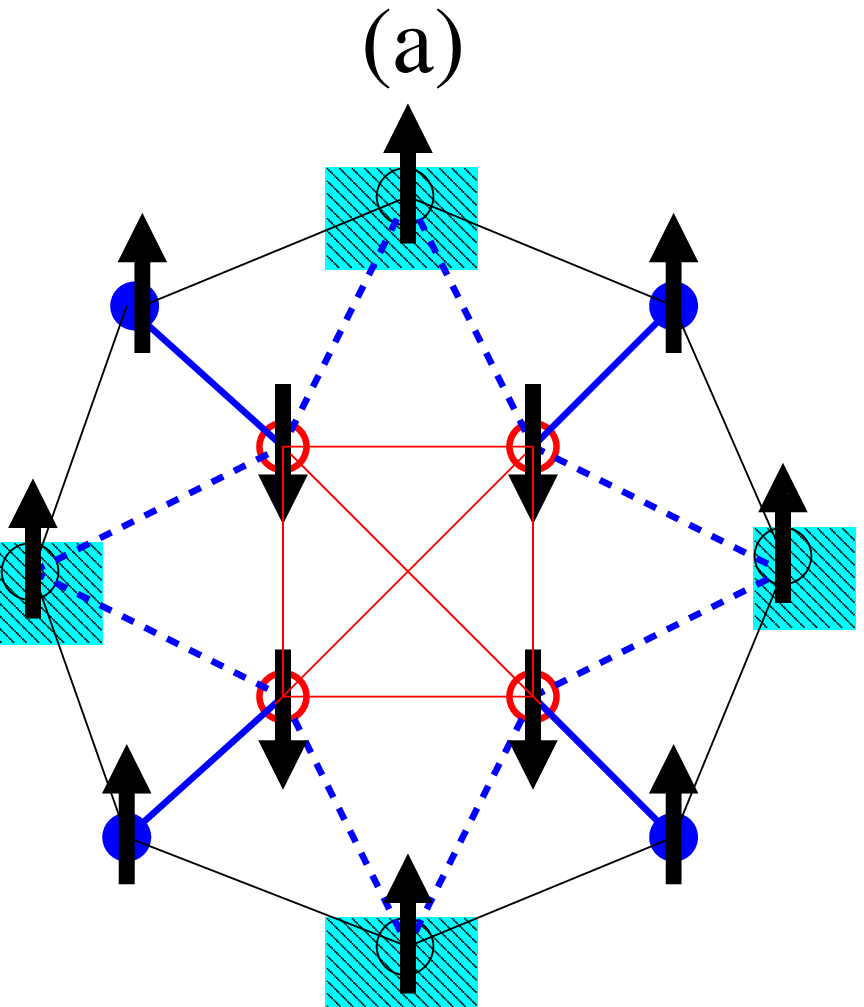}
\hspace{.5cm}
\includegraphics[angle=0,width=0.48\textwidth]{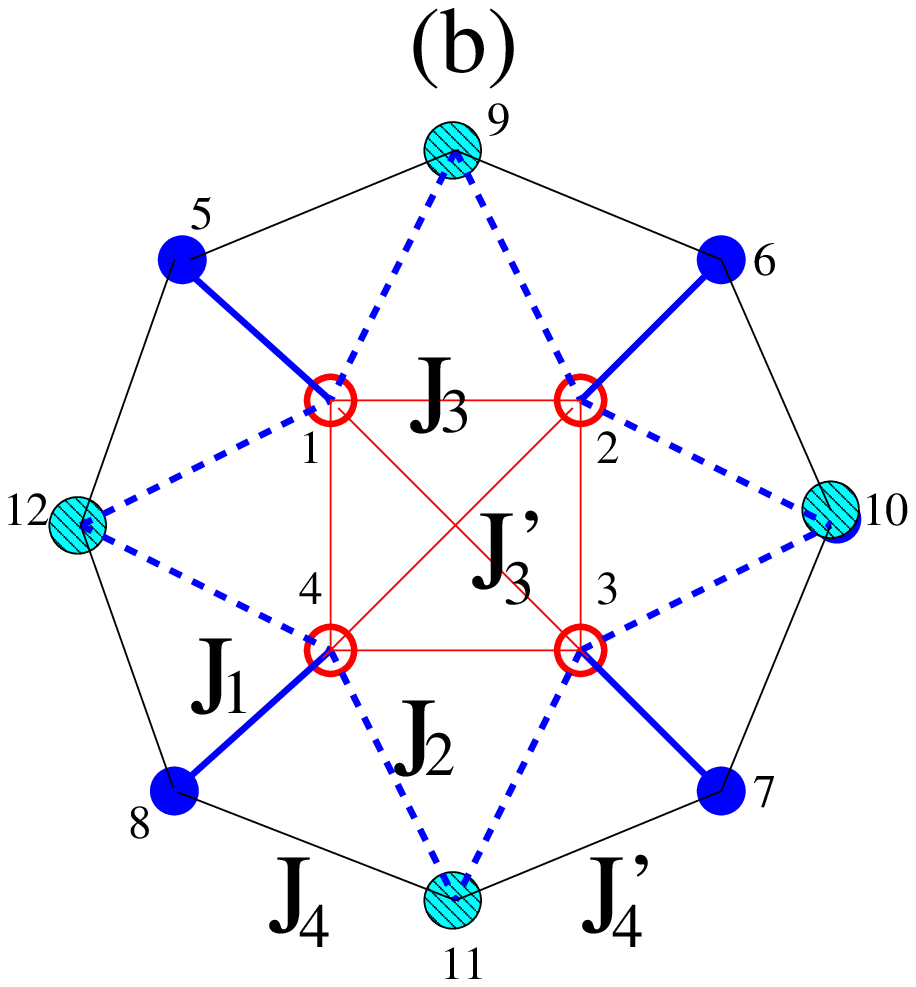}
\caption{(a) Initial spin configuration considered for the fourfold 
symmetric Mn$_{8}$Fe$_4$O$_{12}$(HCOO)$_{16}$(H$_2$O)$_4$ geometry
to determine the most favorable sites for extra electrons. The inequivalent sites
are indicated by the three different symbols. (b) Schematic diagram 
of the exchange constants, $J_1$, $J_2$, $J_3$, $J_3^{\prime}$, $J_4$, and
$J_4^{\prime}$, where $J_3$ and $J_3^{\prime}$ ($J_4$ and $J_4^{\prime}$)
are slightly different from each other because of a difference in 
the bond length by approximately 0.1\AA.~Each spin is labeled from 
1 to 12. The sites $\alpha$ correspond to spins 1-4, the sites 
$\beta$ to spins 5-8, and the sites $\delta$ to spin 9-12.}
\label{fig:spin_config}
\end{figure}

Our initial study showed that the Mn$_8$Fe$_4$ geometry with the total moment
of 24$\mu_B$ has lower energy than that with the total moment of 22$\mu_B$.
The measurement on the experimental Mn$_8$Fe$_4$, however, revealed that
the static magnetic susceptibility for this SMM has a completely different 
temperature dependence from that of the Mn$_{12}$ ($S=10$), which led
to the ground state of $S=2$ for the former.\cite{SCHA94} So following
the method discussed in Refs.~\onlinecite{PARK04-1,KORT01} we examine 
several low-spin states to search for the ground state. In this study, 
we consider collinear spin states only. For a particular total moment, 
there are numerous spin configurations. We thus select a few spin 
configurations to test, considering
that the exchange interactions between the Mn ions and Fe ions through 
the oxygen ligands (i.e. superexchange interactions)\cite{FAZE99} prefer 
antiferromagnetic coupling. We calculate the energies for six selected spin 
configurations with $M_s=2$ and for five chosen configurations with $M_s=4$ 
where $M_s$ is an eigenvalue of $S_z$. Some of the configurations and 
their calculated energies are specified in Table~\ref{table:Mn8Fe4_spin_confs} 
and Fig.~\ref{fig:spin_config}(b). Our calculations show that the ground 
state may have the total moment of 8$\mu_B$ ($M_s=4$) with the spin 
configuration indicated in the sixth row of Table~\ref{table:Mn8Fe4_spin_confs} 
and Fig.~\ref{fig:spin_config}(b). Apparently this does not agree with 
the original experiment\cite{SCHA94} but a very recent experiment on 
[Mn$_8$Fe$_4$O$_{12}$(O$_2$CCH$_2$Cl)$_{16}$(H$_2$O)$_4$] suggested
that the ground state may have a spin of $S=4$.\cite{TSAI04} Notice that 
the CH$_2$Cl molecules in this compound take up the position of the CH$_3$ 
molecules in the experimental Mn$_8$Fe$_4$ compound. In our calculations,
we terminated the carboxyl groups with hydrogen atoms rather than 
CH$_3$ radicals in the Mn$_8$Fe$_4$ geometry. 
Other possible reasons, such as a highly correlated many-spin wavefunction, 
for this discrepancy will be discussed later
in this section and in Sec.~V. 

We discuss the electronic and magnetic properties for the lowest-energy 
structure of the Mn$_8$Fe$_4$ which has the total moment of 8$\mu_B$ and 
the spin configuration specified in Table~\ref{table:Mn8Fe4_spin_confs} 
and Fig.~\ref{fig:spin_config}(b). The energy gap between the majority HOMO and 
the minority (majority) LUMO is about 0.664~eV (0.247~eV). So the 
ground state is magnetically stable. The local moments of the inner Mn 
(the site $\alpha$), outer Mn (the site $\beta$), and outer Fe ions (the site $\delta$)
are found to be 2.61~$\mu_B$, 3.48~$\mu_B$, and -4.16~$\mu_B$, respectively, 
within a sphere of 2.23 Bohr radius. Since the wave functions of Fe and Mn ions
overlap with those of other neighboring atoms, the sphere around 
each ion does not capture a full moment. When we compare the local moments 
with those for the Mn$_{12}$ shown in the second column of 
Table~\ref{table:four_Fe_sub}, we find that only the local moment of 
the outer Fe ion substantially increases. Based on the calculations, 
we speculate that the inner and outer Mn ions and outer Fe ions 
correspond to Mn$^{4+}$ ($S=3/2$), Mn$^{3+}$ ($S=2$), and Fe$^{3+}$ ($S=5/2$), 
respectively. This gives the ground-state spin of 
$S=4 \times 3/2 + 4 \times 2 - 4 \times 5/2 = 4$.
The total MAB is calculated to be 33~K, which is approximately 60\% of 
the MAB for the ordinary Mn$_{12}$, 54~K (Table~\ref{table:Fe_sub_summary}). 
Since the ground state has S$_4$ symmetry, there is no transverse magnetic 
anisotropy, $E=0$. Using the method described in Refs.\onlinecite{BARU02,PARK04-3}, 
we calculate the projected MABs and the local easy axes of the three different 
metal sites. As shown in Table~\ref{table:proj_aniso} (the third column), 
the local magnetic anisotropy of the Mn ions at the sites $\beta$ mainly 
contributes to the total MAB. In contrast to the Mn$_{12}$ geometry, 
the Fe ions at the sites $\delta$ have little contributions to the total MAB. 
This is supported by the calculated local easy axes of the three sites. 
The local easy axis at the site $\alpha$ is nearly in the $x$-$y$ plane. 
The site $\delta$ has a local easy axis 65.8$^{\circ}$ away from 
the $z$ axis (global easy axis), while the site $\beta$ has a local easy
axis only 8.5$^{\circ}$ away from the $z$ axis. As illustrated at the bottom 
of Fig.~\ref{fig:Mn8Fe4}, an Fe$^{3+}$ ion in a distorted octahedral 
environment has exactly half-occupied $e_g$ and $t_{2g}$ $3d$ orbitals, 
which leads to an absence of Jahn-Teller distortion as would be the
case for a Mn$^{4+}$ ion
and a Mn$^{2+}$ ion. Thus the total MAB arises from single-ion anisotropy 
of the four Mn ions at the sites $\beta$ only.

Since the Mn$_8$Fe$_4$ geometry has a very different ground state from
the Mn$_{12}$, it is worthwhile to explore the exchange interactions
between metal ions within a Mn$_8$Fe$_4$ molecule. For this purpose,
we consider collinear spin configurations with $M_s=5$, 6, 7, and 8
in addition to the previously discussed several configurations with 
$M_s=2$, $M_s=4$ and $M_s=12$ (Table~\ref{table:Mn8Fe4_spin_confs}). 
The technique to calculate the exchange interactions using DFT was 
described elsewhere.\cite{PARK04-1,KORT01,PARK04-2} 
As illustrated in Fig.~\ref{fig:spin_config}(b), there are six different exchange
coupling constants within a Mn$_8$Fe$_4$ molecule such as $J_1$, $J_2$, $J_3$,
$J_3^{\prime}$, $J_4$, and $J_4^{\prime}$. Here $J_3$ ($J_4$) 
is slightly different from $J_3^{\prime}$ ($J_4^{\prime}$) because of
a small difference in the bond lengths. To compare with other references, 
we take an average of the two slightly different exchange constants. 
The spin Hamiltonian considered for the calculation of the exchange constants is 
\begin{eqnarray}
{\cal H}_{\mathrm{ex}}&=& E_0 + J_1 \sum_{i=1}^{4} S_i S_{i+4} 
+ J_2 \{ \sum_{i=1}^{3} (S_i + S_{i+1}) S_{i+8} + (S_4 + S_1) S_{12} \} 
+ J_3 \{ \sum_{i=1}^{3} S_i S_{i+1} + S_4 S_1 \} \nonumber \\ 
& & + J_3^{\prime} \sum_{i=1}^{2} S_i S_{i+2}
+ J_4 \{ S_5 S_{12} + \sum_{i=6}^{8} S_{i} S_{i+3} \}
+ J_4^{\prime} \sum_{i=5}^{8} S_{i} S_{i+4} 
\label{eq:ham_ex}
\end{eqnarray}
where $E_0$ is the background energy. All spins are of Ising type and 
they are numbered as illustrated in Fig.~\ref{fig:spin_config}(b). 
From the DFT-calculated energies of the fourteen configurations
(Table~\ref{table:Mn8Fe4_spin_confs}), we extract the seven unknown parameters, 
$E_0$, $J_1$, $J_2$, $J_3$, $J_3^{\prime}$, $J_4$, and $J_4^{\prime}$ 
using a least-square-fit method. Table~\ref{table:Mn8Fe4_exchange} exhibits our 
calculated exchange constants in comparison with experimentally 
estimated values and the exchange constants for the Mn$_{12}$ geometry.
\cite{NOTE01} In contrast to the Mn$_{12}$ geometry, the Mn$_8$Fe$_4$ geometry 
has strong antiferromagnetic interactions between the outer Mn and Fe ions
($J_4$ between the sites $\beta$ and $\delta$), which are comparable to the 
interactions between the inner Mn and outer Mn ions ($J_1$ between the sites 
$\alpha$ and $\beta$ and $J_2$ between the sites $\alpha$ and $\delta$). 
This tendency was also observed in the experimental estimates. 
For a quantitative comparison of the calculated values with experiment,
more refined experiments which are sensitive to the exchange constants
are required. With the calculated exchange constants, one can construct
the isotropic Heisenberg Hamiltonian. The true ground state is obtained
by exact diagonalization of the Heisenberg Hamiltonian.
\cite{PARK04-1,KORT01,PARK04-2} 
The size of the Hilbert space to diagonalize is 207 360 000. We do not
further pursue the ground state of the system by exact diagonalization 
of the large Hamiltonian matrix.
The MAB varies with spin configurations from 16~K to 33~K as shown in
the last column of Table~\ref{table:Mn8Fe4_spin_confs}. This large 
variation was not found for the SMMs Mn$_{12}$ and Mn$_4$ monomer.
\cite{PARK04-1,PARK04-2} We speculate that the reason may be related to 
the fact that for the Mn$_8$Fe$_4$ the considered spin configurations
have energies of the same order of single-electron excitations
(Table~\ref{table:Fe_sub_summary}).

\subsection{Double and single Fe substitutions: Mn$_{10}$Fe$_2$ and Mn$_{11}$Fe}

There are two inequivalent possible ways to replace two Mn ions at the
sites $\delta$ by Fe ions. One way is to place two Fe ions in a twofold 
symmetric fashion ("{\it trans}"). The other is to place Fe ions in a 
unsymmetrized fashion ("{\it cis}"). With the spin configuration shown
in Fig.~\ref{fig:spin_config}(a), we obtain an optimized total moment 
of 22$\mu_B$ for the {\it trans} and {\it cis} geometries. The experiment on 
the SMM (PPh$_4$)$_2$Mn$_{12}$ suggested that the ground-state may have
a total spin of $S=10$.\cite{SOLE03} So we consider several spin 
configurations with $M_s=10$ ({\it trans}) that might have
lower energies than the energy of the configuration with $M_s=11$. 
A good candidate is the configuration obtained by flipping only 
two spins coupled by $J_1$ from the configuration with $M_s=11$ 
(Refer to Figs.~\ref{fig:spin_config}(a) and (b)) We find that 
the configuration with $M_s=10$ has an energy 0.014~eV above 
the energy of the considered configuration with $M_s=11$ 
(Table~\ref{table:one_two_Fe_sub}). 
Thus, the ground state has a total moment of 22$\mu_B$. 
We calculate the electronic and magnetic properties for the ground 
state. For the {\it trans} geometry, the energy gap between the 
majority HOMO and the minority (majority) LUMO is 0.075~eV (0.432~eV). 
The value of $E$ is 0.045~K and the MAB is 31~K 
(Table~\ref{table:Fe_sub_summary}). The calculated projected
anisotropy on the sites $\alpha$, $\beta$, and $\delta$ implies
that only the site $\beta$ contributes to the total MAB 
(Table~\ref{table:proj_aniso}).
For the {\it cis} geometry, the energy gap between the majority HOMO and 
the minority (majority) LUMO is 0.030~eV (0.427~eV). The value of
$E$ is 0.020~K and the MAB is 33~K. The magnetic easy axis is tilted 
by 19$^{\circ}$ from the $z$ axis (Table~\ref{table:Fe_sub_summary}). 
Notice that the value of $E$ for the {\it cis} geometry does not cancel out. 
This contrasts with a negligibly small $E$ value 
calculated for the $n=2$ {\it cis} configuration in a Mn$_{12}$ molecule 
with disordered solvent.\cite{CORN02,PARK04-3} 
The MAB for the {\it cis} geometry agrees with that for the {\it trans} within
numerical uncertainty. This barrier is significantly reduced compared 
to that for Mn$_{12}$ due to an absence of the Jahn-Teller distortion 
in the two Fe${^3+}$ ions and noticeable values of $E$.

When we replace one Mn ion at the site $\delta$ by Fe, with the spin 
configuration shown in Fig.~\ref{fig:spin_config}(a), we obtain an optimized 
total moment of 21$\mu_B$. The experiment on the SMM (PPh$_4$)Mn$_{12}$
suggested that the ground state has a spin of $S=19/2$.\cite{EPPL95} 
We thus consider a few spin configurations with $M_s=19/2$ that might 
have lower energies than that for the considered configuration with 
$M_s=21/2$. A good candidate for this is obtained by flipping only 
two spins coupled by $J_1$ from the starting spin configuration. 
We find that the configuration with $M_s=19/2$ has higher energy 
than the $M_s=21/2$. So the ground state has a total moment of 
21$\mu_B$. The ground state has the energy gap between the majority 
HOMO and the minority (majority) LUMO of 0.043~eV (0.424~eV). 
The total MAB is 41~K and $E=0.03$~K. The magnetic easy axis is 
tilted by 10.7$^{\circ}$ from the $z$ axis. 

\section{DFT calculations: Potassium addition}

\begin{figure}
\includegraphics[angle=0,width=0.8\textwidth]{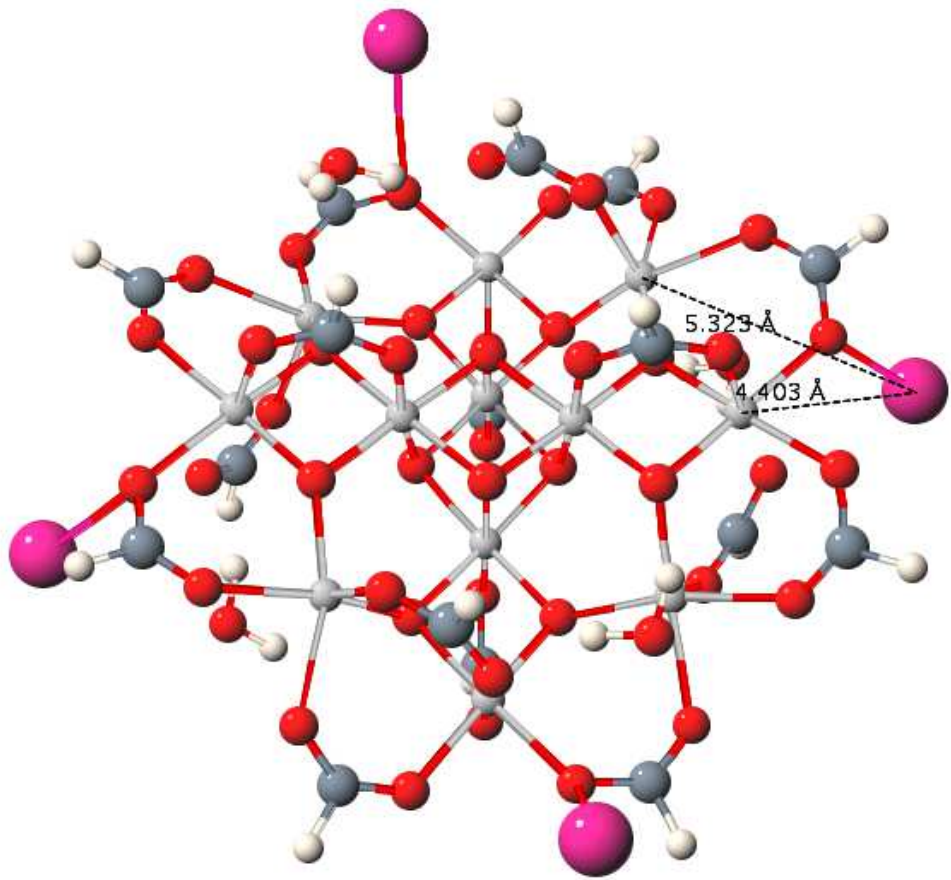}
\caption{Schematic diagram of the fourfold symmetric 
K$_4$[Mn$_{12}$O$_{12}$(HCOO)$_{16}$(H$_2$O)$_4$] geometry,
where two of the K cations are located slightly above the $x$-$y$ plane
and the rest two are slightly below the $x$-$y$ plane. The shading scheme 
is the same as that for Fig.~\ref{fig:Mn8Fe4} except that the largest
spheres are now K cations. The distance between the closest Mn $\beta$ site
($\delta$ site) and K is 4.403 \AA~(5.323 \AA).}
\label{fig:K4}
\end{figure}

\begin{figure}
\includegraphics[angle=0,width=0.8\textwidth]{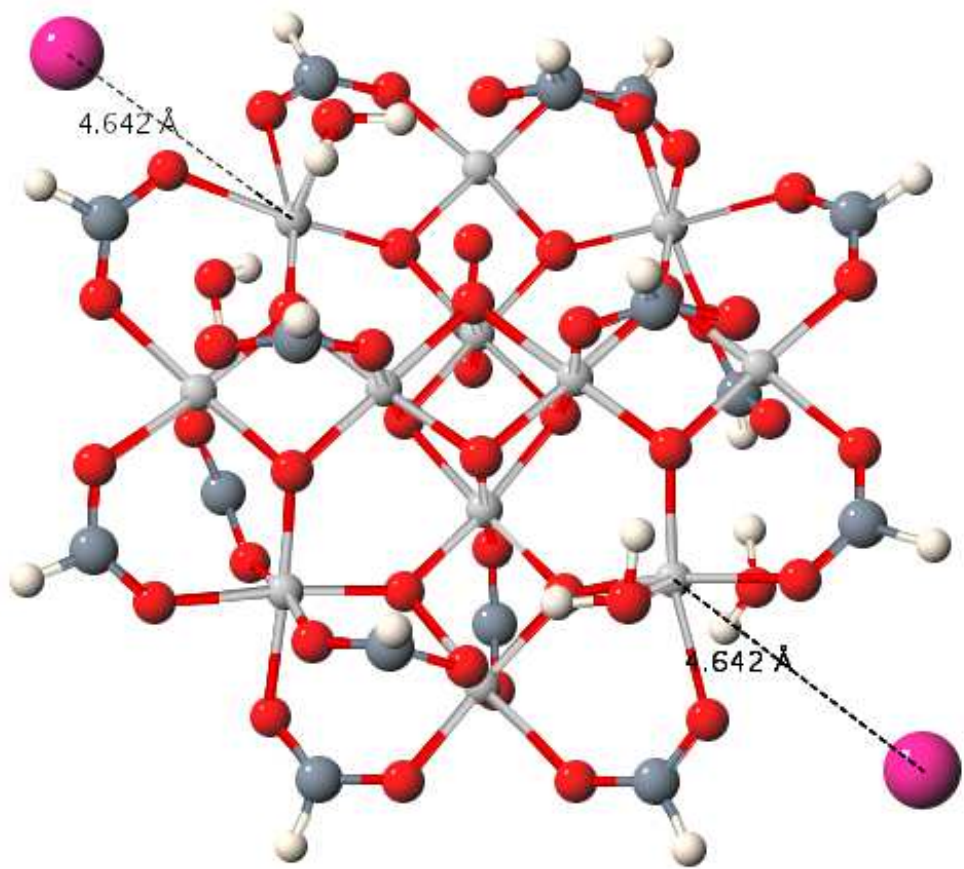}
\caption{Schematic diagram of the {\it twofold} symmetric
K$_2$[Mn$_{12}$O$_{12}$(HCOO)$_{16}$(H$_2$O)$_4$] geometry,
where two K cations are close to the site $\delta$ in the $x$-$y$ plane.
The shading scheme is the same as that for Fig.~\ref{fig:K4}. 
The distance between the closest Mn and K is about 4.642 \AA.}
\label{fig:K2}
\end{figure}

Adding electron donors such as K to the Mn$_{12}$ molecule,
we consider the following two choices for the positions of the donors. 
One is to place the donors at the centers of mass of the acetic acid 
solvent molecules in the Mn$_{12}$ (Fig.~\ref{fig:K4}). In this case, 
the donors are closer to the site $\beta$ than the site $\delta$ by 
0.92~\AA.~The other is to place the donors close to the favorable site 
$\delta$ in the $x$-$y$ plane (Fig.~\ref{fig:K2}). In our calculations,
we use the former (latter) choice for four (two and one) K addition. 
If K atoms were too far away from the Mn$_{12}$ molecule, then the 
unpaired electrons of the K atoms would not be transferred to the 
Mn$_{12}$ molecule. The optimum distance between the K atoms and 
the Mn$_{12}$ molecule is found by relaxation of the geometry.
When two K atoms are added, for direct comparison with experiment,\cite{SOLE03} 
we consider the twofold symmetric Mn$_{12}$ molecule (Fig.~\ref{fig:K2}) 
as well as the S$_4$ symmetric Mn$_{12}$ molecule. The twofold 
symmetric Mn$_{12}$ geometry is obtained from the S$_4$ symmetric Mn$_{12}$
molecule by exchanging formates (HCOO) with neighboring water molecules 
in a twofold symmetric fashion (Fig.~\ref{fig:K2}). All K added geometries
we considered are charge neutral. We use a fine integration mesh and 
full basis sets for all atoms except for K atoms.
We present results for four-, two-, and one-extra electrons 
added to the Mn$_{12}$ molecule.

\subsection{Four K addition: K$_4$Mn$_{12}$}

Four K atoms are added to the Mn$_{12}$ molecule in a S$_4$ symmetric 
fashion shown in Fig.~\ref{fig:K4}). To compare with the experimental
result, similarly to the four-Fe substituted geometry, we consider the
three different spin configurations given in Table~\ref{table:K4_add} 
with three different magnetic moments ($M_s=$2, 4, 12). The three 
considered configurations 
are speculated lowest-energy states for a particular $M_s$. We find that 
the configuration with $M_s=12$ provides the lowest energy among the 
three (Table~\ref{table:K4_add}). In contrast to the Fe substitution, 
when four K atoms are added, the high-spin state with $S=12$ is the 
ground state. Plausible causes for this difference will be discussed 
later in this section and Sec.~V. For the ground state, the local moments 
of the three sites ($\alpha$, $\beta$, and $\delta$) captured within 
a sphere of 2.23 Bohr radius are -2.56~$\mu_B$, 3.79~$\mu_B$, and 
4.16~$\mu_B$, respectively. Comparing these with the local 
moments for the Mn$_{12}$ geometry (Table~\ref{table:K4_add} vs 
Table~\ref{table:four_Fe_sub}) confirms that the spin density of the 
extra electrons are transferred from the K atoms to primarily the 
site $\delta$ in the Mn$_{12}$ molecule even though the site $\beta$ 
is closer to the K cation than the site $\delta$.
This is in accord with the calculated projected density of Mn($3d$) 
orbitals (Fig.~\ref{fig:DOS}).  
As specified in Table~\ref{table:K4_add}, the local moment of the site 
$\beta$ for the $M_s=12$ state is 0.3~$\mu_B$ higher than those 
for the $M_s=2$ and $M_s=4$ states. This indicates that for the 
$M_s=12$ state the spin density of the extra electrons is not 
entirely localized at the site $\delta$ but there is some leakage 
to the site $\beta$. This may contribute to the different
ground state from the Mn$_8$Fe$_4$ geometry. 

We discuss the electronic and magnetic properties of the ground
state. The energy gap for the majority HOMO and the minority (majority) 
LUMO is 1.04~eV (0.62~eV). See Table~\ref{table:K_add_summary}.
The calculated total MAB is 24~K. 
The projected MAB for the site $\delta$ is calculated to be 0.71~K 
and the easy axis of the Mn ion at the site is 51.1$^{\circ}$ 
tilted away from the $z$ axis, which is qualitatively different from 
the projected MAB and the easy axis for the same site in the Mn$_{12}$
geometry. See Table~\ref{table:proj_aniso}. 
This verifies that the absence of a  Jahn-Teller instability at the site 
${\delta}$ reduces the total MAB down to 24~K. This is only 
44\% of the total (2nd-order) MAB for the Mn$_{12}$. We notice that 
the projected MAB of the site $\beta$ is almost half of that of the 
same site for the Mn$_{12}$ or Mn$_8$Fe$_4$ geometry. This might be 
attributed to incomplete localization of the spin density at the 
site $\delta$.

\subsection{Double and single K additions: K$_2$Mn$_{12}$ and KMn$_{12}$}

We first consider the case that two K atoms are located close to the
site $\delta$ where two water molecules are bonded in a twofold
symmetric Mn$_{12}$ molecule. The geometry is referred to as
K$_2$Mn$_{12}$({\it trans})$^{\star}$. 
The optimized total moment is found to be 22~$\mu_B$. Similarly to the
double-Fe substituted geometry, we find that 
the lowest-energy of the $M_s=10$ configurations is higher 
than that for the considered $M_s=11$ configuration as shown in 
Table~\ref{table:one_two_K_add}. So the ground state has a total spin of 
$S=11$. The captured local moments on the Mn ions at the site $\delta$ 
are found to be -4.15~$\mu_B$ within a sphere of 2.23 Bohr radius, 
while those of 
the rest of Mn ions do not change much compared to those for the ordinary
Mn$_{12}$. The energy gap between the majority HOMO and the minority 
(majority) LUMO is 0.74~eV (0.31~eV). The calculated MAB is 31~K and 
$E=0.087$~K (Table~\ref{table:K_add_summary}). 

Now we examine the case when two K atoms are added close to the sites
$\delta$ in the $x$-$y$ plane to the S$_4$ symmetric Mn$_{12}$.
As in the case of the Mn$_{10}$Fe$_2$ geometry, this molecule also has
{\it trans} and {\it cis} geometries. For the {\it cis}, however, the HOMO-LUMO
energy gaps do not open up upon geometry relaxation. Thus we consider 
the {\it trans} geometry only. The energy gap between the majority HOMO 
and the minority (majority) LUMO is 0.82~K (0.29~K). The calculated 
MAB is 34~K and $E=0.076$~K. In comparison with the 
K$_2$Mn$_{12}$({\it trans})$^{\star}$, 
the HOMO-LUMO energy gaps, the total MAB, 
and the $E$ values agree with each other.

When only one K is added close to the site ${\delta}$ in the $x$-$y$ plane, 
the optimized total moment is 21~$\mu_B$. Similarly to the 
single-Fe-substitution case,
we find that the lowest-energy state with $M_s=19/2$ has higher energy than the 
considered configuration with $M_s=21/2$. So the ground state has a total 
spin of $S=21/2$ (Table~\ref{table:one_two_K_add}). The ground state has 
an energy gap between the majority HOMO and the minority 
(majority) LUMO of 0.47~eV (0.036~eV). The calculated MAB is 46~K
and $E=0.033$~K. The magnetic easy axis is tilted by 4.1$^{\circ}$ 
from the $z$ axis (Table~\ref{table:K_add_summary}).

\section{Discussion}

When several Fe ions were substituted for Mn ions at the most favorable 
$\delta$ sites in the Mn$_{12}$ geometry, we found that the total spin changes 
as a function of the number of substituted Fe ions (extra electrons)
(See Table~\ref{table:Fe_sub_summary}). The change of the total spin was 
the most drastic for the case of four Fe substitution. For the Mn$_{12}$, 
Mn$_{11}$Fe, and Mn$_{10}$Fe$_{2}$ geometries, the antiferromagnetic 
exchange interactions between the sites $\beta$ and $\delta$ 
($J_4$ and $J_4^{\prime}$) are much smaller than $J_1$ and $J_2$ 
[Fig.~\ref{fig:spin_config}(b)]. So the ground state has a high spin.
Meanwhile, for the Mn$_{8}$Fe$_{4}$ geometry, the values of $J_4$ and 
$J_4^{\prime}$ are large enough to flip the spins of the inner Mn ions 
and the spins of the Fe ions in comparison to the spin configuration 
of the Mn$_{12}$[Fig.~\ref{fig:spin_config}(a)]. This leads to a 
ground state of a low spin.
Although it is certain that the ground 
state must have a low spin, to determine the relative ordering of the
$S=2$ and $S=4$ states, we need to rely on exact diagonalization of 
the Heisenberg exchange Hamiltonian. The ground state may be a noncollinear 
quantum state (a linear combination of Slater determinants).

\begin{figure}
\includegraphics[angle=0,width=0.8\textwidth]{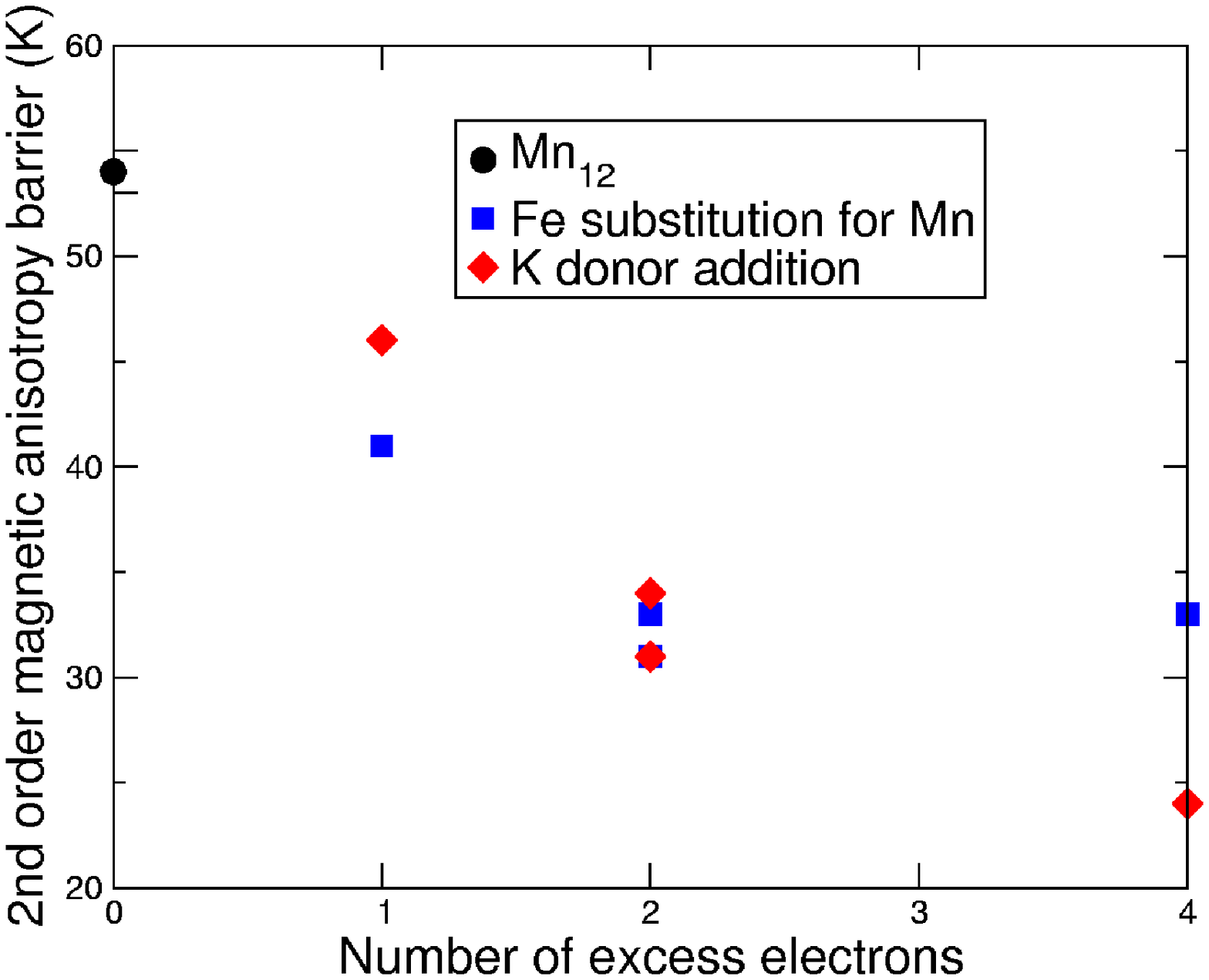}
\caption{Calculated MAB vs number of extra electrons
for the Fe substituted and K added Mn$_{12}$ geometries 
and the Mn$_{12}$ geometry.}
\label{fig:MAB}
\end{figure}

The MAB decreases with increasing number of extra electrons as shown in 
Fig.~\ref{fig:MAB} because substituted Fe ions do not contribute to the 
total MAB due to a lack of Jahn-Teller distortion. Based on rigid band
and intuitive argument, it was expected that the four Fe substitution 
would reduce the MAB more significantly than the MAB for 
two Fe substitution. This, however, was not found in our calculations.
Although the reason is not entirely understood,
it may be related to the result that the MAB varies up to a factor of 2 
depending on the spin configurations given a particular geometry
(Table~\ref{table:Mn8Fe4_spin_confs}). This tendency has 
not been observed for the Mn$_{12}$ and the Mn$_{4}$ monomer where
different spin configurations gave almost the same MAB as that for the ground
state.\cite{PARK04-1,PARK04-2} The major difference between the two cases 
is that for Mn$_{12}$ the spin flip energies are small compared to 
single-electron excitations, while for
the Mn$_8$Fe$_4$ the spin flip energies are comparable or higher than the 
single-electron excitation energy. 
To further understand this large deviation, we calculate the projected 
MABs of the sites $\alpha$, $\beta$, and $\gamma$ for the Mn$_8$Fe$_4$ geometry 
with two different spin configurations ($M_s=4$ and $M_s=12$). We found that
the projected MAB of the site $\alpha$ for $M_s=4$ is smaller than that 
for $M_s=12$ and that the local easy axis of the same site for $M_s=4$ 
is not as close to the $x$-$y$ plane as for the case of $M_s=12$ 
(Table~\ref{table:proj_aniso}). Since the 
projected MAB for the site $\alpha$ counteracts the total barrier or 
the projected MABs for the sites $\beta$ and $\delta$, a smaller barrier 
of this site would probably lead to a larger total barrier. 

To investigate the relationship\cite{PARK04-3} between the values of $E$ 
for the single- and double-Fe substituted geometries ({\it cis} and {\it trans}), 
we use the spin Hamiltonian [Eq.~(\ref{eq:ham_an})] 
for the Mn$_{12}$ and Mn$_{11}$Fe and symmetry. Suppose that the 
spin Hamiltonians for the Mn$_{12}$ and Mn$_{11}$Fe are 
\begin{eqnarray}
{\cal H}[{\mathrm{Mn}}_{12}] &\equiv& \sum_{\mu,\nu=x,y,z} \gamma^{(0)}_{\mu \nu} 
S_{\mu} S_{\nu} \:, \: \: \: \: \: \: {\cal H}[{\mathrm{Mn}}_{11}{\mathrm{Fe}}] 
\equiv \sum_{\mu,\nu=x,y,z} \gamma^{(1)}_{\mu \nu} S_{\mu} S_{\nu} \:,
\label{eq:gamma}
\end{eqnarray}
where diagonalization of the matrices $\gamma^{(0)}$ ($\gamma^{(1)}$) would
provide the values of $D$ and $E$. Then the spin Hamiltonian for the double-Fe 
substituted geometry is written, in terms of ${\cal H}$[Mn$_{12}$] 
and ${\cal H}$[Mn$_{11}$Fe], as
\begin{eqnarray}
{\cal H}[{\mathrm{Mn}}_{10} {\mathrm{Fe}}_2] &=& \sum_{\mu,\nu=x,y,z} 
\gamma^{(1)}_{\mu \nu} [ S_{\mu} S_{\nu} + R(S_{\mu}) R(S_{\nu}) ] 
- \sum_{\mu,\nu=x,y,z} \gamma^{(0)}_{\mu \nu} S_{\mu} S_{\nu} \\
 & \equiv & \sum_{\mu, \nu=x,y,z} \gamma^{(2t),(2c)}_{\mu \nu} S_{\mu} S_{\nu} \\
\gamma^{(2t)}_{xy}&=&2\gamma^{(1)}_{xy}, \: \: \: \gamma^{(2t)}_{xz}=0, \: \: \:
\gamma^{(2t)}_{yz}=0, \label{eq:2t} \\
\gamma^{(2c)}_{xy}&=&0, \: \: \: \gamma^{(2c)}_{xz}=\gamma^{(1)}_{xz}-\gamma^{(1)}_{yz},
\: \: \: \gamma^{(2c)}_{yz}=\gamma^{(1)}_{xz}+\gamma^{(1)}_{yz}
\label{eq:2c}
\end{eqnarray}
where $R$ represents a proper symmetry operation and $\gamma^{(2t)}$ and 
$\gamma^{(2c)}$ stand for the magnetic anisotropy matrices for the {\it trans} 
and {\it cis} geometries, respectively. Here the magnetic anisotropy matrix 
$\gamma$ can be calculated within DFT.\cite{PEDE99} Notice that a value of 
$E$ may be determined by all of the off-diagonal elements of the 
matrix $\gamma^{(1)}$, while the tilting angle of the magnetic easy axis is 
influenced only by their $xz$ and $yz$ components. For the single-Fe 
substituted geometry the value of $E$ is obtained by the $xy$, $yz$, and $xz$ 
components of the matrix $\gamma^{(1)}$. The substantially larger 
tilting angle of the easy axis implies that the values of $\gamma_{xz}^{(1)}$ and 
$\gamma_{yz}^{(1)}$ are not as small as those calculated for the Mn$_{12}$ with solvent 
disorder.\cite{CORN02,PARK04-3} For the {\it trans} geometry, due to symmetry, 
$\gamma_{xz}^{(2t)}=0$ and $\gamma_{yz}^{(2t)}=0$ . Symmetry allows the nonzero off-diagonal 
element $\gamma_{xy}^{(2t)}$ to depend on $\gamma_{xy}^{(1)}$ only [Eq.~(\ref{eq:2t})]. 
So the value of $E$ for the {\it trans} geometry is not twice as large as the value of $E$
for the single-Fe substituted geometry. For the {\it cis} geometry, although 
the value of $\gamma^{(2c)}_{xy}$ vanishes due to symmetry, there are nonzero
values of $\gamma^{(2c)}_{xz}$ and $\gamma^{(2c)}_{yz}$. Since $\gamma^{(2c)}_{xz}$ 
and $\gamma^{(2c)}_{yz}$ are not small, the easy-axis tilting and the value of $E$ 
are substantial for the {\it cis} geometry (Table~\ref{table:Fe_sub_summary}). 
The values of $E$ calculated from the relaxed {\it trans} and {\it cis} geometries 
agree well with those obtained from Eqs.~(\ref{eq:2t}) and (\ref{eq:2c}). 
The value of $E$ for the {\it cis} is comparable to that for the {\it trans} 
or single-Fe substituted geometry, while the value of $E$ for the $n=2$ {\it cis} 
in the Mn$_{12}$ with solvent disorder was an order of magnitude smaller
that that for the $n=1$ or $n=2$ {\it trans}. 

When several K atoms were added to the Mn$_{12}$ geometry, we found that
the magnetic structure and properties are primarily the same as those
for Fe substitution except for a few points which will be described
(Table~\ref{table:K_add_summary} vs Table~\ref{table:Fe_sub_summary}).
In both cases, the ground state has the same total spin for each number of
extra electrons, except for four extra electrons. Unlike four Fe 
substitution, the four-K added geometry has the ground state of a high spin, 
$S=12$. Similar to the case for Fe substitution, the MAB decreases with 
increasing 
number of added K atoms. Except for four-extra electrons, the K added 
geometries have the similar MABs to the Fe substituted geometries. The 
relationship between the values of $E$ for the single- and double-Fe 
substituted geometries also holds for the single- and double-K added 
geometries. The values of $E$ and the tilting angles of the easy axes 
are quite different for the two cases. 
We speculate that the difference between the two cases arises from incomplete 
localization of the spin density of the extra electrons on the sites $\delta$ 
for K addition. Another difference is that when one- or two-extra electrons
are added, for K addition the energy gap between the majority HOMO and 
minority LUMO is an order of magnitude larger than that for Fe substitution.
The calculated total MAB and the local easy axis for the sites $\delta$ 
for both the single-K added and single-Fe substituted geometries, 
suggest that the geometries for one-extra electron have features in 
accord with the 
experimentally observed fast-relaxing species in a variety of 
low-symmetry SMMs Mn$_{12}$.\cite{WERN99,SUN99,AUBI01,TAKE02,SUZU03} 

Although the case of the three-extra electrons was not considered in
our DFT calculations, it is possible to calculate the values of $D$
and $E$ using the matrices $\gamma^{(0)}$ and $\gamma^{(1)}$ 
[Eq.~(\ref{eq:gamma})] and symmetry. It is found that for triple-Fe 
substitution $D=-0.20$~K and $E=0.016$~K,
while for triple-K addition $D=-0.29$~K and $E=0.032$~K 
(Tables~\ref{table:Fe_sub_summary} and \ref{table:K_add_summary}). 
For completeness, the values of $D$ and $E$ for the K$_2$Mn$_{12}$
({\it cis}) is also calculated using Eq.~(\ref{eq:gamma}) and symmetry: 
$D=-0.38$~K and $E=0.003$~K. 
The value of $E$ for the K$_2$Mn$_{12}$({\it cis}) is an order
of magnitude smaller than that for the Mn$_{10}$Fe$_2$({\it cis}).
In the event that such a system is experimentally realized, 
the calculated values of $D$ and $E$ can be compared to experiment. 

\section{Conclusion}

To examine the effect of extra electrons on the electronic and magnetic 
structure and the magnetic anisotropy, we considered, within DFT, 
two classes of molecules with extra electrons in the ordinary 
Mn$_{12}$ geometry: 
Fe substitution for Mn and K addition as electron donors. Four Fe
substitution implied that the sites $\delta$ are energetically favorable
for Fe ions, which agreed with experiment. Our calculations showed that
the MAB decreases with increasing number of extra electrons, which is 
caused by absence of Jahn-Teller distortion at the sites where the extra
electrons are localized. This tendency agreed with experiment and was
confirmed by the calculated projected anisotropy on the sites.
The total spin increases with number of extra electrons, except that
the Mn$_8$Fe$_4$ geometry has the low-spin ground state due to the 
enhanced exchange interactions between the Fe and outer Mn ions. 
The S$_4$ symmetry-breaking geometries such as the one- and two-extra 
electron geometries have $E$ values that are at least by a factor of 3 larger 
than those associated solvent disorder. These larger $E$ values will 
significantly expedite the tunneling 
and eventually reduce the total MAB. Additionally, the tilting angles of 
the easy axes are an order of magnitude larger than those for the Mn$_{12}$ 
with solvent disorder, which will provide large internal transverse fields 
that will further expedite tunneling. 

\section*{Acknowledgments}
The authors are grateful to D.~N. Hendrickson, E.~M. Rumberger, and H.-L. Tsai 
for providing information on the published experiment on 
[Mn$_8$Fe$_4$O$_{12}$(O$_2$CCH$_3$)$_{16}$(H$_2$O)$_4$].
The authors were supported in part by ONR and the DoD HPC CHSSI program.

\clearpage

\begin{table}
\begin{center}
\caption{Total fixed moments (2$M_s$), calculated local moments 
(sphere radii of 2.23 Bohr) associated with the three metal sites
($\alpha$, $\beta$, and $\delta$ shown in Fig.~\ref{fig:Mn8Fe4}), 
$\mu_{\alpha}$, $\mu_{\beta}$, and $\mu_{\delta}$, energy differences 
from the geometry $\delta$ with $M_s=12$, $\Delta E$, and whether 
the Fermi filling is satisfied, for the ordinary Mn$_{12}$ geometry 
and for the three initial tetra-Fe substituted geometries 
($\alpha$, $\beta$, and $\delta$), 
[Mn$_8$Fe$_4$O$_{12}$(HCOO)$_{16}$(H$_2$O)$_4$], with 
the spin configuration of Fig.~\ref{fig:spin_config}(a). In the geometries 
$\alpha$, $\beta$, and $\delta$, four Fe ions are located at the four 
sites $\beta$, $\delta$, and $\alpha$ shown in Fig.~\ref{fig:Mn8Fe4}, 
respectively. 
For example, Fig.~\ref{fig:Mn8Fe4} shows the geometry $\delta$. }
\label{table:four_Fe_sub}
\begin{ruledtabular}
\begin{tabular}{c|c||c|c|c||c|c|c}
  & Mn$_{12}$ & $\alpha$ & $\beta$ & $\delta$ & $\alpha$ & $\beta$ & $\delta$ \\ \hline
$M_s$ & 10 & 12 & 12 & 12 & 11 & 11 & 11 \\ \hline
$\mu_{\alpha}$ & -2.58 & -1.55 & -2.59 & -2.54 & -1.84 & -2.59 & -2.57  \\ \hline
$\mu_{\beta}$ & 3.62 & 3.60 & 3.94 & 3.59 & 3.51 & 3.74 & 3.45  \\ \hline
$\mu_{\delta}$ &  3.55 & 3.52 & 3.64 & 3.95 & 3.48 & 3.55 & 3.83  \\ \hline
$\Delta E$ (eV)& & 1.119 & 1.186 & 0  & 1.826 & 1.106 & 0.157 
\\ \hline
Fermi filling & Yes & Yes & No & Yes & No & Yes & No  \\
\end{tabular}
\end{ruledtabular}
\end{center}
\end{table}

\begin{table}
\begin{center}
\caption{Total moments ($2M_s$), down spins (the rest of spins are in an up state
and each spin is numbered following Fig.~\ref{fig:spin_config}(b)), Ising energy 
expressions [Eq.~(\ref{eq:ham_ex})], the energy differences from the lowest-energy 
spin configuration ($M_s=4$) $\Delta E$, the MABs for the fourteen collinear spin 
configurations in the tetra-Fe substituted geometry shown in Fig.~\ref{fig:Mn8Fe4}.
Here $M_s=2$ (V-2) denotes a spin configuration with $M_s=2$ which is different
from the other three configurations with $M_s=2$.
The exchange coupling constants $J_1$, $J_2$, $J_3$, $J_3^{\prime}$,
$J_4$, and $J_4^{\prime}$, are defined in Fig.~\ref{fig:spin_config}(b).}
\label{table:Mn8Fe4_spin_confs}
\begin{ruledtabular}
\begin{tabular}{c|c|c|c|c}
$M_s$ & down spins & Ising energy & $\Delta E$ (eV) & MAB (K) \\ \hline
2  & 2,4,5,6,7,8 & $E_0 -9 J_3 + 4.5 J_3^{\prime} - 20 J_4 - 20 J_4^{\prime}$
& 0.2764 & 26 \\ \hline
2 (II) & 1,3,4,6,8,10 & $E_0 - 6 J_1 - 15 J_2$ - 10 $J_4$ + 10 $J_4^{\prime}$ 
& 0.4016 & \\ \hline 
2 (IV) & 1,8,9,10,11 & $E_0-15J_2-20J_4^{\prime}$ & 0.2747 & 29 \\ \hline
2 (V-2) & 5,6,7,11,12 & $E_0-6J_1+9J_3+4.5J_3^{\prime}-10J_4-10J_4^{\prime}$ 
& 0.3941 & 22 \\ \hline  
4 & 9,10,11,12 & $E_0+12J_1-30J_2 + 9J_3+4.5J_3^{\prime}-20 J_4-20J_4^{\prime}$ 
& 0 & 33 \\ \hline
4 (I) & 1,2,3,4,6,8 & $E_0-30J_2+9J_3+4.5J_3^{\prime}$ & 0.2808 & 22 \\ \hline
4 (II-1) & 1,3,6,11,12 & $E_0-6J_1-9J_3+4.5J_3^{\prime}-10J_4-10J_4^{\prime}$
& 0.4131 & 21 \\ \hline
4 (II-2) & 2,3,6,11,12 & $E_0+6J_1-15J_2-4.5J_3^{\prime}-10J_4-10J_4^{\prime}$
& 0.3262 & 27 \\ \hline
4 (III) & 1,6,7,8,12 & $E_0-12J_1+15J_2-20J_4$ & 0.5265 & 16 \\ \hline
5 & 8,9,10,11 & $E_0+6J_1 -15 J_2 + 9 J_3 +4.5J_3^{\prime}- 20 J_4^{\prime}$ 
& 0.3312 & 32 \\ \hline
6 & 1,10,11,12 & $E_0+6J_1 -15 J_2 -10 J_4-10J_4^{\prime}$ & 0.3407 & 27 \\ \hline
7 & 1,2,4,7,8 & $E_0-6 J_1 -15 J_2$ & 0.4240 & 22 \\ \hline
8 & 5,6,7,8 & $E_0-12 J_1 +30 J_2 + 9J_3+4.5J_3^{\prime}-20 J_4-20J_4^{\prime}$ 
& 0.5334 & 24 \\ \hline
12 & 1,2,3,4 & $E_0-12 J_1 -30 J_2 + 9J_3+4.5J_3^{\prime}+20 J_4+20J_4^{\prime}$ 
& 0.6013 & 19 \\
\end{tabular}
\end{ruledtabular}
\end{center}
\end{table}

\begin{table}
\begin{center}
\caption{Calculated total spins, the second-order magnetic anisotropy parameters
($D$ and $E$), easy-axis tilting angles $\theta$ (angles between the local 
easy axes and the $z$ axis), MABs, the energy gaps between majority HOMO and 
minority (majority) LUMO for the Mn$_{12}$ and the Fe substituted geometries. 
Except for Mn$_9$Fe$_3$ the lowest-energy spin configuration has been 
used for each geometry. For Mn$_9$Fe$_3$ results are determined from
Eq.~(\ref{eq:gamma}) and the use of symmetry.}
\label{table:Fe_sub_summary}
\begin{ruledtabular}
\begin{tabular}{c|c|c|c|c|c|c}
 & Mn$_{12}$ & Mn$_{11}$Fe & Mn$_{10}$Fe$_2$({\it trans}) &
    Mn$_{10}$Fe$_2$({\it cis}) & Mn$_9$Fe$_3$ & Mn$_8$Fe$_4$ \\ \hline
   $S$ & 10 & 21/2 & 11 & 11 & & 4 \\ \hline
   $D$ (K) &-0.54 & -0.42 & -0.29 & -0.33 & -0.20 & -2.10 \\ \hline
   $E$ (K) &0.0  & 0.030 & 0.045 & 0.020 & 0.016 & 0.0 \\ \hline
   $\theta$ (deg) & 0.0 & 10.7 & 0.0 & 19.0 & & 0.0 \\ \hline
   MAB (K) & 54 & 41 & 31 & 33 & & 33 \\ \hline
   maj H - min L (eV) & 0.804 & 0.043 & 0.075 & 0.030 & & 0.664 \\ \hline
   maj H - maj L (eV) & 0.438 & 0.424 & 0.432 & 0.427 & & 0.247 \\
\end{tabular}
\end{ruledtabular}
\end{center}
\end{table}

\begin{table}
\begin{center}
\caption{Calculated projected (local) MABs $\Delta_i$ and tilting angles
of the local easy axes $\varphi_i$ for the sites, $\alpha$, $\beta$, and $\delta$ 
(Fig.~\ref{fig:Mn8Fe4}), in the Mn$_{12}$ ($S=10$), Mn$_8$Fe$_4$ ($S=4$), 
Mn$_8$Fe$_4$ ($S=12$), Mn$_{10}$Fe$_2$ ($S=11$), and K$_4$Mn$_{12}$ ($S=12$)
geometries.}
\label{table:proj_aniso}
\begin{ruledtabular}
\begin{tabular}{c|c|c|c|c|c}
 & Mn$_{12}$($S=10$) & Mn$_8$Fe$_4$ ($S=4$) & Mn$_8$Fe$_4$ ($S=12$) 
 & Mn$_{10}$Fe$_2$ ($S=11$) & K$_4$Mn$_{12}$ ($S=12$) \\ \hline
$\Delta_{\alpha}$ (K) & 0.87 & 0.21 & 0.96 & 1.16 & 0.85  \\ \hline
$\Delta_{\beta}$ (K) & 9.7  & 9.5  & 9.8 & 10.4 & 5.5 \\ \hline
$\Delta_{\delta}$ (K) & 9.3  & 0.82 & 0.41 & 0.95 & 0.71  \\ \hline
$\varphi_{\alpha}$ (degree) & 83.8 & 58.9 & 87 & 83 & 83.4 \\ \hline
$\varphi_{\beta}$ (degree) & 9.8 & 8.5 & 8.3 & 9.4 & 9.1 \\ \hline
$\varphi_{\delta}$ (degree) & 36.2 & 65.8 & 61.5 & 63 & 51.1 \\ 
\end{tabular}
\end{ruledtabular}
\end{center}
\end{table}

\begin{table}
\begin{center}
\caption{DFT-determined and experimental exchange constants for the tetra-Fe 
substituted geometry shown in Fig.~\ref{fig:Mn8Fe4} and the Mn$_{12}$ geometry.
In this study, a fine mesh and full basis sets were used for both the Mn$_8$Fe$_4$ 
and Mn$_{12}$ geometries. For the Mn$_8$Fe$_4$, the fourteen spin configurations 
were used (Table~\ref{table:Mn8Fe4_spin_confs}), while for the Mn$_{12}$, 
the eleven spin configurations were used (Ref.[\onlinecite{PARK04-1}]). 
For comparison with other published results, a DFT-determined value of $J_3$ 
($J_4$) represents an average between the values of $J_3$ and $J_3^{\prime}$ 
($J_4$ and $J_4^{\prime}$) shown in Fig.~\ref{fig:spin_config}(b). 
The exchange constants for the Mn$_{12}$ shown here are slightly different from 
those reported in Ref.~[\onlinecite{PARK04-1}] because in the previous study 
smaller basis sets and fewer exchange constants were used ($J_3=J_3^{\prime}$
and $J_4=J_4^{\prime}$.}
\label{table:Mn8Fe4_exchange}
\begin{ruledtabular}
\begin{tabular}{c|c|c|c|c}
 & Mn$_8$Fe$_4$ (DFT) & Mn$_8$Fe$_4$ (Exp[Ref.~\onlinecite{SCHA94}])
 & Mn$_{12}$ (DFT) & Mn$_{12}$ (Exp[Ref.~\onlinecite{REGN02}]) \\ \hline
$S$ & 4 & 2 & 10 & 10 \\ \hline
 $J_1$ & 119 & 432 & 140 & 119 \\ \hline
 $J_2$ & 150 & 397 & 117 & 118 \\ \hline
 $J_3$ & 7.2 & 173 & 7.2 & -8  \\ \hline
 $J_4$ & 122 & 187 & 24 & 23 \\ 
\end{tabular}
\end{ruledtabular}
\end{center}
\end{table}

\begin{table}
\begin{center}
\caption{Total moments (2$M_s$), down spins, the energy differences from 
the $M_s=21/2$ ($M_s=11$) geometry, and whether Fermi filling 
is satisfied, for the different spin configurations in the Mn$_{11}$Fe 
(Mn$_{10}$Fe$_2$ {\it trans}) geometry. For the Mn$_{11}$Fe geometry, an Fe ion
occupies site 12 shown in Fig.~\ref{fig:spin_config}(b). For the 
Mn$_{10}$Fe$_2$ geometry, two Fe ions occupy sites 9 and 11 shown
in Fig.~\ref{fig:spin_config}(b). The spin configurations for the 
$M_s=21/2$ and $M_s=11$ geometries are the same as Fig.~\ref{fig:spin_config}(a).
The configurations for the $M_s=19/2$ and $M_s=10$ are obtained by exchanging 
spin 1 with spin 5 in Fig.~\ref{fig:spin_config}(b).}
\label{table:one_two_Fe_sub}
\begin{ruledtabular}
\begin{tabular}{|c|c|c|c|c|}
 & \multicolumn{2}{c|}{Mn$_{11}$Fe} & \multicolumn{2}{c|}{Mn$_{10}$Fe$_2$({\it trans})} 
\\ \hline
$M_s$ & 19/2 & 21/2 & 10 & 11 \\ \hline
down spins & 2,3,4,5 & 1,2,3,4 & 2,3,4,5 & 1,2,3,4 \\ \hline
$\Delta E$ (eV) & 0.035 & 0 & 0.014 & 0 \\ \hline
Fermi filling & Yes & Yes & Yes & Yes \\  
\end{tabular}
\end{ruledtabular}
\end{center}
\end{table}

\begin{table}
\begin{center}
\caption{Total moments ($2M_s$), down spins, local moments captured 
around the three sites, $\alpha$, $\beta$, and $\delta$ 
(Fig.~\ref{fig:Mn8Fe4}), the energy differences from the lowest-energy 
spin configuration, whether Fermi filling is satisfied, and MABs, 
for the three different spin configurations in the four-K added Mn$_{12}$ 
geometry, K$_4$[Mn$_{12}$O$_{12}$(HCOO)$_{16}$(H$_2$O)$_4$] 
shown in Fig.~\ref{fig:K4}. For $M_s=2$, there are two slightly different 
local moments because of the twofold symmetry of the spin configuration. 
For $M_s=4$ and $M_s=12$, the spin configurations are fourfold symmetric.}
\label{table:K4_add}
\begin{ruledtabular}
\begin{tabular}{c|c|c|c}
$M_s$ & 2 & 4 & 12 \\ \hline
down spins & 2,4,5,6,7,8 & 9,10,11,12 & 1,2,3,4 \\ \hline
$\mu_{\alpha}$ ($\mu_B$) & 2.78, 2.52 & 2.61 & -2.56 \\ \hline
$\mu_{\beta}$ ($\mu_B$) & 3.44, 3.48 & 3.49 & 3.79 \\ \hline
$\mu_{\delta}$ ($\mu_B$) & -4.20, -4.21  & -4.16 & 4.16 \\ \hline
$\Delta E$ (eV) & 0.576 & 0.359 & 0 \\ \hline
Fermi filling & Yes & Yes & Yes \\ \hline
MAB (K) & 20 & 32 & 24 \\ 
\end{tabular}
\end{ruledtabular}
\end{center}
\end{table}

\begin{table}
\begin{center}
\caption{Calculated total spins, the second-order magnetic anisotropy parameters,
easy-axis tilting angles, the MABs, the energy gaps between majority
HOMO and minority (majority) LUMO for the K added Mn$_{12}$ geometries.
For K$_2$Mn$_{12}$({\it cis}) and K$_3$Mn$_{12}$ results are determined 
from Eq.~(\ref{eq:gamma}) and the use of symmetry.}
\label{table:K_add_summary}
\begin{ruledtabular}
\begin{tabular}{c|c|c|c|c|c|c}
 & KMn$_{12}$ & K$_2$Mn$_{12}$({\it trans}) & 
 K$_2$Mn$_{12}$({\it trans})$^{\ast}$
& K$_2$Mn$_{12}$({\it cis}) & K$_3$Mn$_{12}$ 
& K$_4$Mn$_{12}$ \\ \hline
$S$ & 21/2 & 11 & 11  & & & 12\\ \hline
$D$ (K) & -0.46 & -0.36 & -0.35 & -0.38 & -0.29 & -0.16 \\ \hline
$E$ (K) & 0.033 & 0.076 & 0.087 & 0.003 & 0.032 & 0.0 \\ \hline
$\theta$ (degree) & 4.1 & 0.0  & 0.0 & & & 0.0 \\ \hline
MAB (K) & 45.7 & 34 &  31  & & & 24 \\ \hline
maj H - min L (eV) & 0.47 & 0.82 & 0.74 & & & 1.04  \\ \hline
maj H - maj L (eV) & 0.036 & 0.29 & 0.31 & & & 0.62  \\
\end{tabular}
\end{ruledtabular}
\end{center}
\end{table}

\begin{table}
\begin{center}
\caption{Total moments, down spins, local moments, the energy differences
from the lowest-energy spin configuration, and whether Fermi filling
is satisfied, for two different spin configurations in the double K-
and single-K added Mn$_{12}$ geometries. In the {\it trans} geometry marked by $\ast$, 
the twofold symmetric Mn$_{12}$ geometry (Fig.\ref{fig:K2}) was used.}
\label{table:one_two_K_add}
\begin{ruledtabular}
\begin{tabular}{|c|c|c|c|c|}
 & \multicolumn{2}{c|}{K[Mn$_{12}$O$_{12}$]} & \multicolumn{2}{c|}
{K$_2$[Mn$_{12}$O$_{12}$]$^{\ast}$ ({\it trans})} \\ \hline
$M_s$ & 19/2 & 21/2 & 10 & 11 \\ \hline
down spins & 2,3,4,5 & 1,2,3,4 & 2,3,4,5 & 1,2,3,4 \\ \hline
$\Delta E$ (eV) & 0.0823 & 0 & 0.3225 & 0 \\ \hline
Fermi filling & Yes & Yes & Yes & Yes \\ 
\end{tabular}
\end{ruledtabular}
\end{center}
\end{table}

\begin{thebibliography}{10}
\bibitem{ISOT}
Considering the natural abundances of the isotopes of H, C, O, and Mn atoms,
we find that 37\% of Mn$_{12}$ molecules in a single crystal may have
one or two isotopes.
\end{thebibliography}
\end{document}